\documentclass[twocolumn,showpacs,preprintnumbers,amsmath,amssymb]{revtex4}

\usepackage{graphicx}
\usepackage{dcolumn}
\usepackage{bm}
\newcommand{\be}{\begin{equation}}
\newcommand{\ee}{\end{equation}}
\newcommand{\bea}{\begin{eqnarray}}
\newcommand{\eea}{\end{eqnarray}}
\newcommand{\ba}{\begin{array}}
\newcommand{\ea}{\end{array}}
\newcommand{\bi}{\begin{itemize}}
\newcommand{\ei}{\end{itemize}}

\renewcommand{\vec}[1]{\mbox{\boldmath $#1 \!\!$ \unboldmath}}

\begin{document}

\title{The nucleon resonances in the $J/\psi \to p\bar{p}\eta'$ decay}

\author{Xu Cao$^{1,2,3}${\footnote{Electronic address: caoxu@impcas.ac.cn}} }
\author{Ju-Jun Xie$^{1,2}${\footnote{Electronic address: xiejujun@impcas.ac.cn}}}

\affiliation{$^1$Institute of Modern Physics, Chinese Academy of Sciences, Lanzhou 730000, China\\
$^2$State Key Laboratory of Theoretical Physics, Institute of Theoretical Physics, Chinese
Academy of Sciences, Beijing 100190, China\\
$^3$Kavli Institute for Theoretical Physics China (KITPC), Chinese Academy of Sciences, Beijing 100190, China}

\begin{abstract}

  We are aiming to study the $J/\psi \to p\bar{p}\eta'$ decay in an isobar model and the effective Lagrangian approach. After a careful exploration of the contributions of the $S_{11}(1535)$, $P_{11}(1710)$, $P_{13}(1900)$, $S_{11}(2090)$ and $P_{11}(2100)$ resonances,  we conclude that either a subthreshold resonance or a broad $P$-wave state in the near threshold range seems to be indispensable to describe present data of the $\pi N \to \eta'N$. Furthermore, at least one broad resonance above $\eta'N$ threshold is preferred. With this detailed analysis, we could give the invariant mass spectrum and Dalitz plot of the $J/\psi \to p\bar{p}\eta'$ decay for the purpose of assisting the future detailed partial wave analysis. It is found that the $J/\psi \to p\bar{p}\eta'$ data are useful for disentangling the above or below threshold resonant contribution, though it still further needs the differential cross section data of $\pi N \to \eta'N$ to realize some of the resonant and the non-resonant contribution. Our results are enlightening for the $\eta'N$ production mechanism and the properties of the nucleon resonances with the mass around 2.0~GeV.

\end{abstract}
\maketitle{}

\section{Introduction} \label{sec:intro}

In recent years, plenty of information on nucleon resonances~\cite{pdg2012} has been obtained by a wealth of phenomenological studies on numerous data of the $\pi N$, $\gamma N$, $eN$ reactions~\cite{BoCh11,BoCh12,BoCh08,BoCh12qtet,Feuster98,Penner2002,Shklyar04J,Shklyar05ome,Shklyar07eta,Shklyar12eta,Shklyar05lam,caoKSigma,Manley1992,Manley2012,Arndt2012,Doring11,Doring12,Huang12,LeePhysRe,KamanoCC,Kamano2pi} and $pN$ collisions~\cite{BCLiuPRL06,JJXiePLB07,JJXiePRC08,JJXiePRC10,JJXiePRC11,JJXiePRC13,caoCPL2008,caoetap2008,caophi2009,caoCPC2009,caotwopi,caoNPA2011,caoIJMPA}. However, despite a great deal of theoretical and experimental efforts, our knowledge of nucleon resonances around 2.0~GeV is still scarce because of the presence of many resonances and opening channels in this energy region. Alternatively, the hadronic decay channels of heavy quarkonium have attracted much attention due to their advantage in extracting empirical information of resonances with isospin $1/2$. In this area, a lot of progress have been made on the study of the decay of the charmonium states, e.g. $J/\psi$, $\psi(3686)$, $\psi(3770)$ and $\chi_{cJ}$ states by BES and CLEO collaborations~\cite{BESRoper06,Besnstar2,Besnstar3,Besnstar4,Besnstar5,Besnstar6,Besnstar7,BESPRD06,BESPRL2013,CLEO2005,CLEO2007}. In particular, it is advanced by the wide implementation of the tools of partial wave analysis (PWA)~\cite{ZouEPJA2003,ZouPRC2003,ZouEPJA2005} to the tremendous (up to billion) events accumulated with the BESIII detector at the BEPCII facility. In these fruitful PWA works which mainly concentrate on the $N\bar{N}\pi$ channels, not only the peak of known $N^*$ resonances are directly observed, but also the evidence of several new resonances with higher mass are found~\cite{BESRoper06,BESPRL2013,Besnstar5}, e.g. $N^*(2040)$ with $J^P = 3/2^+$ found in $J/\psi \to p \bar{p} \pi^0$~\cite{Besnstar5} and $N^*(2300)$ with $J^P = 1/2^+$ and $N^*(2570)$ with $J^P = 5/2^-$ appeared in $\psi(3686) \to p \bar{p} \pi^0$~\cite{BESPRL2013}.

It is indispensable to explore the decay modes with final mesons other than $\pi$-meson in order to search for the missing states coupling weakly to $\pi$-meson. Unfortunately, up to date we know little about the coupling of the $\eta'$-, $\omega$-, and $\phi$-mesons to nucleon resonances~\cite{Torres13} and the interaction of these mesons with the nucleon~\cite{OsetPLB11}. In past decades, the production of these mesons in the $\gamma N$ and $p N$ reactions have been widely investigated, mainly motivated by the increasing data measured by CLAS, CBELSA and COSY groups~\cite{JJXiePLB07,JJXiePRC08,JJXiePRC10,JJXiePRC11,JJXiePRC13,caoCPL2008,caoetap2008,caophi2009,Nakappeta1,Nakappeta2,HuangPRC13}. But the results are quite inconclusive for the moment. It is still not firmly established which resonance(s) play the important role in these reactions, and it is still controversial whether the sub-threshold resonances have essential contributions. In order to resolve these ambiguities, it is natural to deliver our sight into the strangeness decay of charmonium states, e.g.  $N\bar{N}\eta'$, $N\bar{N}\phi$, and the associate strangeness decay channels $N\bar{\Lambda}K$ and $N\bar{\Sigma}K$~\cite{Besnstar4}.

The invariant mass spectrums of the $J/\psi \to N\bar{N}\eta'$ decay cover the energy range from $m_N + m_{\eta'} \simeq 1.90$~GeV to $m_{J/\psi} - m_N \simeq 2.16$~GeV, where the debatable $P_{13}(1900)$ state~\cite{BoCh08,Penner2002} and the long-sought third $S_{11}$ and $P_{11}$ states at about 2100 MeV~\cite{BoCh08,BoCh11,BoCh12,BoCh12qtet,Feuster98,Penner2002} are expected to be present. The $P_{13}(1900)$ state, which is unfavored by diquark models, is considered at the early stage of Giessen model~\cite{Penner2002,Shklyar04J,Shklyar05lam,Shklyar05ome,Shklyar07eta} and KSU survey~\cite{Manley1992,Manley2012}. The Bonn-Gatchina partial wave analysis find its evidence in the $K\Sigma$ photoproduction data only recently~\cite{BoCh12,BoCh08}, but the latest GWU analysis do not include it as before~\cite{Arndt2012}. The existence of the third $S_{11}$ and $P_{11}$ states could shed light on the spin quartet of nucleon resonances, which is disputed in classical diquark models~\cite{BoCh12qtet}. This topic is interesting also because it could shed light on the nature and the internal structure of relevant nucleon resonances which may have large $s\bar{s}$ component~\cite{Torres13}. It can also serve as a guideline to the future detailed PWA in view of the current scarce information on these resonances.

In the mentioning decay channels, the possible background contribution - the nucleon pole - has been calculated to be negligible, except in the $J/\psi \to N\bar{N}\pi$~\cite{WHLiangJPG02,WHLiangEPJA04,BarnesPRD10}, as anticipated by the suppression from the large off-shell effect. The meson-pole Feynman diagrams, e.g. $J/\psi \to M\eta' \to (p\bar{p})\eta'$, could also be ignored because of the smallness of the relevant coupling~\cite{Okubo1984}. So the main contribution should come from the nucleon resonances. In an isobar model including several possible resonances coupling strongly to $N\phi$, the $J/\psi \to p\bar{p}\phi$ has been studied and useful hints are given for the future data analysis~\cite{DaiPRD12}. In this paper, we will give a full study of the $J/\psi \to p\bar{p}\eta'$ decay based on our present understanding of the resonances with mass around 2.0~GeV with the model parameters constrained by the data of $\pi N \to \eta'N$ channel.

In the next section, the construction of the model and the mathematical framework is presented in detail. Section~\ref{sec:result} is continued to demonstrate the calculated results, followed by a short summary in section~\ref{sec:summary}.

\section{Ingredients and Formalism} \label{sec:formalism}

\begin{figure}
  \begin{center}
{\includegraphics*[width=8cm]{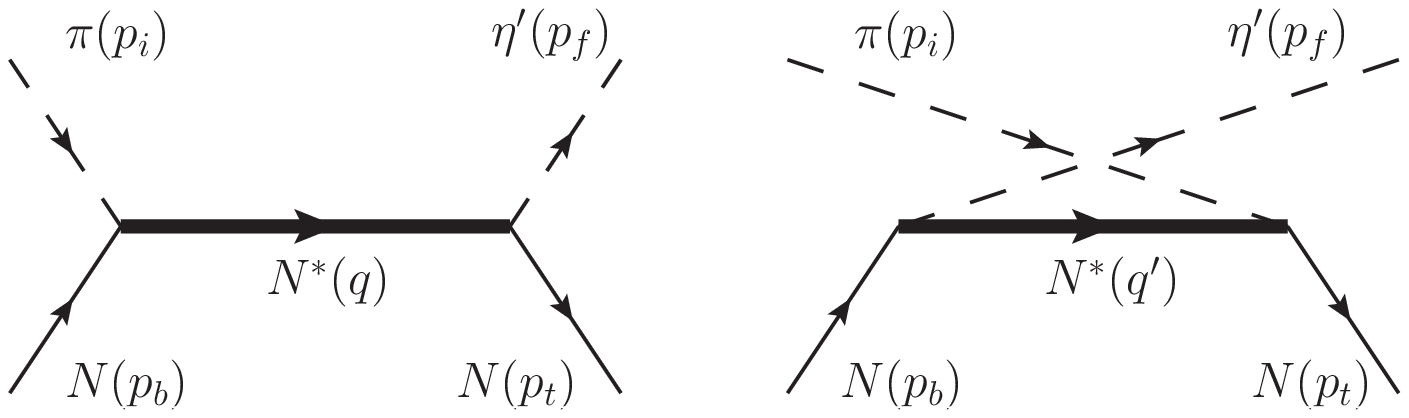}}
       \caption{
  The $s$- and $u$-channel Feynman diagram for the $\pi N \to \eta'N$ reaction in the isobar model.
      \label{feydiagr:SU}}
  \end{center}
\end{figure}

\begin{figure}
  \begin{center}
{\includegraphics*[width=8cm]{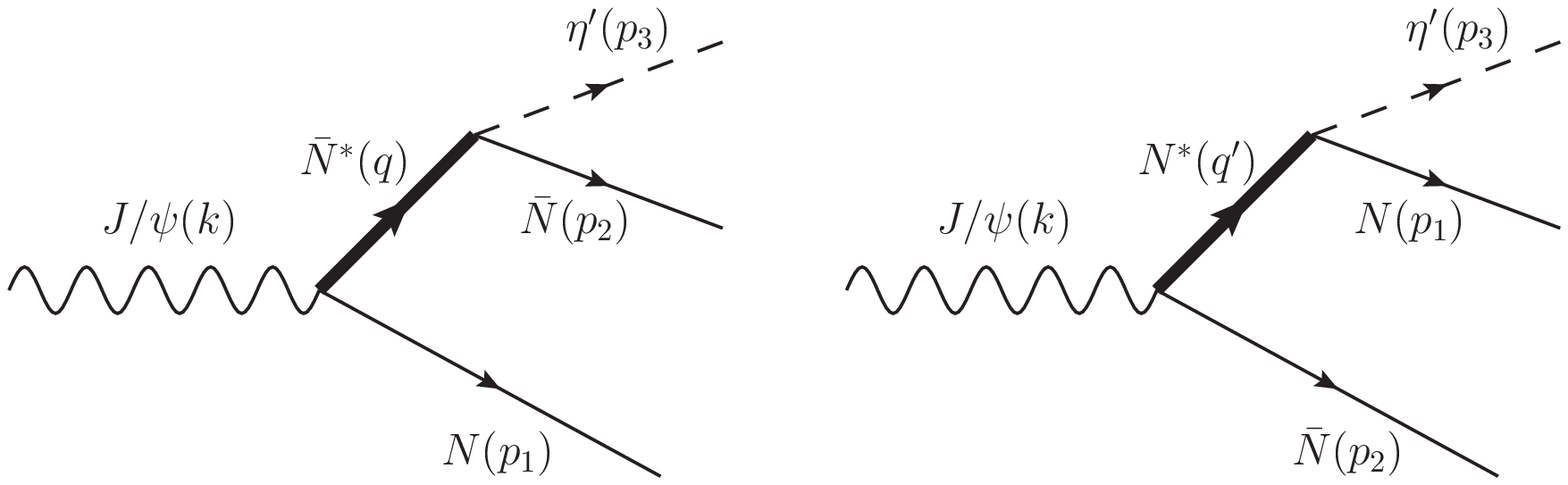}}
       \caption{
  The Feynman diagram for the $J/\psi \to p\bar{p}\eta'$ decay  in the isobar model.
      \label{feydiagr:Jpsi}}
  \end{center}
\end{figure}

We use the isobar model with the assumption of nucleon resonances dominance in the paper. We use the available data of the $\pi N \to \eta'N$ reaction to determine the unknown coupling constants of $\eta' NN^*$ vertices. The $s$- and $u$-channel process as depicted in Fig.~\ref{feydiagr:SU} are included in the model, but the $t$-channel contribution is not considered because we do not find any meson coupling strongly to $\pi\eta'$. The nucleon pole~\cite{HuangPRC13} is calculated to be very small so we disregard it, too. As pointed out above, the invariant mass of $\eta' N$ and $\eta' \bar{N}$ in the $J/\psi \to p\bar{p}\eta'$ decay cover the energies up to $2.16$~GeV, so herein we consider the center of mass (c.m.) energy range from threshold to $2.5$~GeV in the $\pi N \to \eta'N$ reaction in order to better constraint the coupling constants of the $\eta' NN^*$ vertices. The Feynman diagrams of the $J/\psi \to p\bar{p}\eta'$ decay in the model are shown in Fig.~\ref{feydiagr:Jpsi}. We use the experimental branching ratios (BR) of $J/\psi \to p\bar{p}\eta$ and $J/\psi \to p\bar{p}\pi^0$ to extract the coupling constants of $J/\psi NN^*$, whose Feynman diagrams are similar to Fig.~\ref{feydiagr:Jpsi}, but replacing the final meson by the $\pi$- and $\eta$-meson, respectively.

Before continuing to the formalism, we first explain the strategy of selecting the nucleon resonances in the model. We only include the $S$- and $P$-wave states because the energy we consider here is not far away from threshold, and also the evidence of higher spin resonances coupling to $\eta' N$ is little~\cite{pdg2012}. A further reference worthy to be mentioned is that the PWA in the $J/\psi \to p\bar{p}\eta$ do not find any signal of higher partial waves~\cite{Besnstar2}. In addition, we consider the known states with relatively big partial decay widths to strangeness or associate strangeness channels, because these states are expected to couple strongly to the $\eta' N$ channel. The $P_{13}(1900)$ state, locating very close to the $\eta' N$ threshold, seems to have big couplings to $\eta N$, $K\Sigma$, $K\Lambda$~\cite{pdg2012} and $N\phi$~\cite{DaiPRD12}. However its decay width to $\eta' N$ is critically suppressed by the very small phase space, which is probably the reason of its obscurity in the presnt PWA analysis. Just above the threshold is situated the $S_{11}(2090)$ and $P_{11}(2100)$, which may have big couplings to $\eta' N$ but rank only two and one star, respectively in the compilation of Particle Data Group (PDG)~\cite{pdg2012}. The $S_{11}(2090)$ is labeled as $S_{11}(1895)$ in the latest PDG with the recommended Breit-Wigner mass to be around 2090~MeV. We include the sub-threshold resonance $S_{11}(1535)$ because it may have big couplings to $\eta' N$, as found  in $pN \to pN\eta'$ reaction~\cite{caoetap2008}. In fact, the constitute of this resonance may have significant $s \bar{s}$ element~\cite{AnSSG09,AnPRC11}, resulting in its big couplings to $K\Lambda$~\cite{BCLiuPRL06} and $N\phi$~\cite{JJXiePRC08,caophi2009,JShiPRD11}. However, the combination of several $S$- and $P$-wave  resonances above $\eta' N$ threshold could give reasonable description to the data of $pN \to pN\eta'$~\cite{Nakappeta1,Nakappeta2,HuangPRC13}. One of the purposes in this paper is to explore whether it is possible to discriminate these two $\eta'$ production mechanisms in the $J/\psi \to p\bar{p}\eta'$ decay. Another sub-threshold resonance $P_{11}(1710)$ is considered in the model because it has relatively big decay width to the $\eta N$. Other sub-threshold resonances, i.e. the $S_{11}(1650)$ and $P_{11}(1880)$ state~\cite{pdg2012,BoCh12} is not considered because they can not be distinguished from the nearby $S_{11}(1535)$ and $P_{11}(1710)$ resonances by the current $\eta'$ production data. So in this paper we include $S_{11}(1535)$, $P_{11}(1710)$, $P_{13}(1900)$, $S_{11}(2090)$ and $P_{11}(2100)$, labeled as $1 \sim 5$ with the increasing masses. As can be seen in the PDG~\cite{pdg2012}, the widths of the last three resonances have big discrepancy between different models. Herein we adopt the treatment in Ref.~\cite{DaiPRD12} and use three widths of each resonance, determined by three different models~\cite{pdg2012}, labeled respectively as (a), (b) and (c) from narrow one to wide one. We try to use the combinations of these widths to fit the data with less model parameters. In the following text, we label different fitting strategies as \emph{nr.ixjykz...}, which means its fitting with $n$ resonances with the (x), (y) and (z) etc. widths for $i$, $j$ and $k$ etc. resonances, respectively. For example, the 3r.2-3b4c is fitting with 3 resonances, which is $P_{11}(1710)$, $P_{13}(1900)$ with its (b) width, and $S_{11}(2090)$ with its (c) width.

\begin{table*}[t]
  \begin{center}
  \begin{tabular}{cccrccccccc}
 \hline
Label & $N^*$ & mass(MeV) & $\Gamma_{tot}$(MeV) & $BR_{\pi N}$(\%) & $g^2_{\pi NR}$ & $\Gamma_{J/\psi N N^*}$(~keV) & $g^2_{\psi NR}(10^{-5})$ \\
 \hline
  1 & $S_{11}(1535)$ & 1530 &    137.5  & 45.0  & 0.47   & 9.94$\times10^{-2}$ & 0.652  \\
 \hline
  2 & $P_{11}(1710)$ & 1695 &     75.0  & 15.0  & 1.08   & 1.23$\times10^{-2}$ & 0.316  \\
 \hline
  3 & $P_{13}(1900)$ & 1900 &(a) 180.0  &  5.5  & 1.13   & 1.23$\times10^{-2}$ & 2.422  \\
    &                &      &(b) 250.0  & 10.0  & 2.85   & 1.23$\times10^{-2}$ & 1.475  \\
    &                &      &(c) 498.0  & 26.0  & 14.7   & 1.23$\times10^{-2}$ & 0.774  \\
 \hline
  4 & $S_{11}(2090)$ & 2090 &(a)~ 95.0   &  9.0  & 0.041  & 1.23$\times10^{-2}$ & 22.05  \\
    &                &      &(b) 350.0  & 18.0  & 0.305  & 1.23$\times10^{-2}$ & 7.521   \\
    &                &      &(c) 414.0  & 10.0  & 0.200  & 1.23$\times10^{-2}$ & 13.20   \\
 \hline
  5 & $P_{11}(2100)$ & 2100 &(a) 113.0  & 15.0  & 0.564  & 1.23$\times10^{-2}$ & 1.362   \\
    &                &      &(b) 200.0  & 10.0  & 0.666  & 1.23$\times10^{-2}$ & 2.290   \\
    &                &      &(c) 260.0  & 12.0  & 1.040  & 1.23$\times10^{-2}$ & 2.031   \\
 \hline
 \end{tabular}
  \end{center}
  \caption{The parameters of nucleon resonances used in the calculation. The Breit-Wigner masses, widths and branching ratios (BR) are quoted from the central values of the Particle Data Group (PDG)~\cite{pdg2012}.
  \label{Tab:ccNstar}}
\end{table*}

In order to evaluate the Feynman diagrams in Fig.~\ref{feydiagr:SU} and Fig.~\ref{feydiagr:Jpsi}, we construct the effective Lagrangians with the covariant $L$-$S$ (obital-spin) coupling scheme~\cite{caotwopi,caoNPA2011,ZouEPJA2003,ZouEPJA2005,ZouPRC2003}. The couplings of pseudoscalar meson ($M = \vec\tau \cdot \vec{\pi}$, $\eta$ or $\eta'$) to $S_{11}$, $P_{11}$ and $P_{13}$ resonances ($R$) are:
\bea
{\cal L}^{1/2^\pm}_{MNR} = g_{MNR} \bar{N} \Gamma^{\pm} M R_{\mu} + h.c. \quad,
\label{eq:p11}
\eea
\bea
{\cal L}^{3/2^+}_{MNR} = i \frac{g_{MNR}}{m_R} \bar{N} \partial^{\mu} M R_{\mu} + h.c. \quad,
\label{eq:p13}
\eea
with $\Gamma^{-} = 1$ and $\Gamma^{+} = i\gamma_5$ for $R = S_{11}$ and $P_{11}$, respectively. The couplings of $J/\psi$ to resonances are:
\bea
{\cal L}^{1/2^\pm}_{\psi NR}=g_{\psi NR}\bar{N} \Gamma^{\pm}_{\mu} \epsilon^{\mu}(\vec{p}_{\psi},s_{\psi}) R + h.c. \quad,
\label{eq:psip13}
\eea
\bea
{\cal L}^{3/2^+}_{\psi N R} & = & i g_{\psi NR} \bar{N} \gamma_5 \epsilon^{\mu}(\vec{p}_{\psi},s_{\psi}) R_{\mu} + h.c. \quad,
\label{eq:psiNNstar1900}
\eea
with $\Gamma^-_{\mu} = i \gamma_{5}\sigma_{\mu\nu}p^{\nu}_{\psi}/m_{N}$ and $\Gamma^+_{\mu} = \gamma_{\mu} $ for $R = S_{11}$ and $P_{11}$, respectively. It should be noted that the $J/\psi$ produced in $e^+e^-$ collisions is transversely polarized so the polarization vector $\epsilon_{\mu}(\vec{p},,s_{\psi})$ is satisfying,
\bea
\sum_{s_{\psi}=\pm1}\epsilon_{\mu}(\vec{p},,s_{\psi}) \epsilon^{*}_{\nu}(\vec{p},s_{\psi})= \delta_{\mu\nu}(\delta_{\mu1}+\delta_{\mu2}) \quad.
\label{eq:polzsum}
\eea

The intermediate resonances are multiplied by off-shell form factors to suppress the contribution of high momentum:
\bea
F_{R}(q^2)=\frac{\Lambda_R^{4}}{\Lambda_R^{4} + (q^2-m^2_R)^2} \quad,
\label{eq:ffactor}
\eea
with $\Lambda_R$ and $q$ being, respectively, the cut-off parameter and four-momentum of the resonances. The $\Lambda_R = 1.1$ and 2.0 are used in the $s$- and $u$-channel in the $\pi N \to \eta'N$ reaction, respectively. In the $J/\psi$ decay channels, the $\Lambda_R = 1.8$ and 2.3 are adopted for the resonances below and above threshold, respectively. The propagators of the resonances with total spin $J = 1/2$ and $3/3$ are:
\bea
G^{1/2}_{R}(q)=\frac{ -i(\not \! q \pm m_{R})}{q^2-m^2_{R}+im_{R}\Gamma_{R}} \quad,
\label{eq:prophalf}
\eea
\bea
G^{3/2}_{R}(q)=G^{1/2}_{R}(q) G_{\mu\nu}(q) \quad,
\label{eq:prop32}
\eea
\bea
G_{\mu \nu}(q) &=& - g_{\mu \nu} + \frac{1}{3} \gamma_\mu \gamma_\nu \pm \frac{1}{3 m_R}( \gamma_\mu q_\nu - \gamma_\nu q_\mu) \nonumber \\ &+& \frac{2}{3 m^2_R} q_\mu q_\nu \quad,
\label{eq:prop32mn}
\eea
where $\pm$ are for the particles and antiparticles, respectively.

The partial decay widths of nucleon resonances could be calculated by the above Lagrangins, e.g. Eq.~(\ref{eq:p11}) and Eq.~(\ref{eq:p13}) as following:
\bea
\Gamma_{R \to N M} &=& \frac{g^2_{MNR} (E_N \pm m_N)p^{cm}_{N}}{4\pi m_R} \Gamma_{J} \quad,
 \label{eq:widthRNpi}
\eea
with $\Gamma_{1/2} = 3$ and $\Gamma_{3/2} = (p^{cm}_N/m_R)^2$. The $\pm$ is for $S_{11}(P_{13})$ and $P_{11}$, respectively. The energy of the nucleon in the rest frame of resonances is $E_N=\sqrt{(p^{cm}_{N})^2+m^2_N}$, with the corresponding momentum
\bea
p^{cm}_{N}=\sqrt{\frac{(m^2_{R}-(m_N+m_{M})^2) (m^2_{R}-(m_N-m_{M})^2)}{4m^2_{R}}} \quad
\label{eq:widthpcm}
\eea
So the coupling constants of $MNR$ vertex, as listed in Tab.~\ref{Tab:ccNstar}, could be determined by the experimental decay widths of $R \to NM$ in the compilation of PDG~\cite{pdg2012}. Because the parameters of $P_{13}(1900)$, $S_{11}(2090)$ and $P_{11}(2100)$ have large uncertainties, their widths are adopted from three PWA groups, see PDG~\cite{pdg2012} for details. The branch decay ratio of $S_{11}(1535)$ and $P_{11}(1710)$ to $\eta N$ channel are 53\% and 6.2\%, respectively, with the resulting values $g^2_{\eta N N^*(1535)} = 4.31$ and $g^2_{\eta N N^*(1710)} = 3.14$.

After some algebraic manipulation, the separate amplitudes of the $\pi^\pm N \to \eta'N$ is written as,
\bea
{\cal M}_R &=& \sqrt{2} g_{\eta' N R} g_{\pi N R} \times \nonumber \\ && [e^{i\phi_s} F_{R}(q) \bar{u}(p_t) \Gamma_{\eta' NR} G_{R}(q) \Gamma_{\pi NR} u(p_b) \nonumber \\ &+& e^{i\phi_u} F_{R}(q') \bar{u}(p_t) \Gamma_{\pi NR} G_{R}(q') \Gamma_{\eta' NR} u(p_b)] \quad
\label{eq:piNamps}
\eea
with $q = p_b + p_i$ and $q' = p_b - p_f$ for $s$- and $u$-channel, respectively. Here the interaction vertices $\Gamma_{\eta' NR}$ and $\Gamma_{\pi NR}$ could be read directly from Eq.~(\ref{eq:p11}) and Eq.~(\ref{eq:p13}) as $1$, $i\gamma_5$ and $i p^\mu_{f,i}/m_R$ for $R = S_{11}$, $P_{11}$ and $P_{13}$ respectively. The total amplitudes is the coherent sum of all resonances with the relative phase setting as free parameters to be determined by the data.


The $J/\psi \to N\bar{N}M (M=\pi$ or $\eta')$ decay amplitude for each resonance could be written as,
\bea
{\cal M}_R &=& \delta_M e^{i\phi_s} g_{MNR} g_{\psi NR} \times \nonumber \\ && [F_{R}(q) \bar{v}(p_2) \Gamma_{MNR} G_{R}(q) \Gamma_{\psi NR} \epsilon^{\mu}(\vec{p}_{\psi},s_{\psi}) \bar{u}(p_1) \nonumber \\ &+& F_{R}(q') \bar{u}(p_1) \Gamma_{MNR} G_{R}(q') \Gamma_{\psi NR} \epsilon^{\mu}(\vec{p}_{\psi},s_{\psi}) v(p_2)] \quad \quad
\label{eq:Jpsiamps}
\eea
with the isospin factor $\delta_M = 1$ or $\sqrt{2}$ for neutral or charged final mesons, respectively. The vertices $\Gamma_{\psi NR}$ could be read directly from Eq.~(\ref{eq:psip13}) and Eq.~(\ref{eq:psiNNstar1900}) as $i \gamma_{5}\sigma_{\mu\nu}p^{\nu}_{\psi}/m_{N}$, $\gamma_{\mu}$ and $i\gamma_5$ for $R = S_{11}$, $P_{11}$ and $P_{13}$ respectively.

We use the same method to determine the $g_{\psi NR}$ as in Ref.~\cite{DaiPRD12} and update the values with the amplitudes in Eq.~(\ref{eq:Jpsiamps}). In BES's PWA~\cite{Besnstar2}, the branch ratio of the $J/\psi \to p\bar{p}\eta$ is determined to be $(1.91 \pm 0.02\pm 0.17) \times 10^{-3}$, among which the $S_{11}(1535)$ contributes $(56\pm 15)\%$. So the $g_{\psi NN^*(1535)}$ could be well determined, taking advantage of the above value of the $g_{\eta NN^*(1535)}$. Combining with the known $g_{\pi NN^*(1535)}$, we can predict the fraction contribution of the $S_{11}(1535)$ in the $J/\psi \to p\bar{p}\pi^0$ is $1.83 \times 10^{-2}$~keV, compatible with the branching fraction $(0.92 \sim 2.10) \times 10^{-4}$ in BES's PWA~\cite{Besnstar5}. However, the contribution of other resonances in $J/\psi \to N\bar{N}M$ is not well disentangled by PWA and they depend on the model parameters and the selected sets of resonances~\cite{Besnstar5}. The total branch ratio of the $J/\psi \to p\bar{n}\pi^- + c.c.$ is consistent with expectation from the $J/\psi \to p\bar{p}\pi^0$ with the isospin relation but the fraction contribution of separate resonances is not extracted yet~\cite{BESRoper06}. Based on the uncertainties given by the BES's analysis~\cite{Besnstar5}, we can safely assume that the fraction contribution of the $P_{11}(1710)$, $P_{13}(1900)$, $S_{11}(2090)$ and $P_{11}(2100)$ are all 10\% in $J/\psi \to p\bar{p}\pi^0$, then the $g_{\psi NR}$ could be determined, as tabulated in Tab.~\ref{Tab:ccNstar}. These values are also good starting point to calculate other $J/\psi$ hadronic decay channels, e.g. the $J/\psi \to N\bar{N}\pi\pi$ channel.

\section{Numerical results} \label{sec:result}

\begin{table}[t]
  \begin{center}
  \begin{tabular}{cccccc}
 \hline
  Label/$\chi^2$   & $N^*$ & width & $g_{\eta'NR}$ & $\phi_{s}(^\circ)$ & $\phi_{u}(^\circ)$  \\
 \hline
  1r.2-/2.77  & 1710   & ---   &  $12.97\pm 0.88$  &  0.0$^*$          &  $123.8 \pm 8.7$  \\
 \hline
  2r.2-4c & 1710   & ---   &  $9.95\pm 1.85$   &  0.0$^*$          &  $128.9 \pm 18.3$  \\
   2.32   & 2090   & (c)   &  $0.71\pm 0.26$   &  $0.0\pm 274.6$   &  $156.5 \pm 213.2$  \\
 \hline
  3r.1-2-5b & 1535   & ---   &  $3.02\pm 1.48$   &  0.0$^*$          &  $138.8 \pm 138.1$  \\
            & 1710   & ---   &  $10.83\pm 2.37$  &  $0.0\pm 55.4$    &  $21.2 \pm 58.5$  \\
   2.36     & 2100   & (b)   &  $2.18\pm 1.39$   &  $43.0\pm 84.8$   &  $248.0 \pm 119.0$  \\
 \hline
  3r.2-3b4c & 1710   & ---   &  $6.96\pm 2.25$   &  0.0$^*$          &  $277.0 \pm 77.9$  \\
            & 1900   & (b)   &  $12.37\pm 3.26$  &  $139.0\pm 89.5$  &  $1.5 \pm 300.5$  \\
   2.10     & 2090   & (c)   &  $1.73\pm 1.03$   &  $62.4\pm 64.5$   &  $297.7 \pm 299.7$  \\
 \hline
  4r.1-2-3c4a & 1535   & ---   &  $9.48\pm 4.17$   &  0.0$^*$           &  $0.9 \pm 324.9$  \\
              & 1710   & ---   &  $9.14 \pm 1.22$  &  $0.0 \pm 352.8$   &  $4.8 \pm 242.8$   \\
              & 1900   & (c)   &  $10.06\pm 2.93$  &  $269.8\pm 106.8$  &  $359.9 \pm 352.1$ \\
   2.02       & 2090   & (a)   &  $0.93\pm 0.98$   &  $138.3\pm 80.0$   &  $1.7 \pm 208.1$  \\
 \hline
  5r.1-2-3c4c5c & 1535   & ---  &  $5.68\pm 6.38$    &  0.0$^*$          &  $1.6 \pm 37.8$  \\
                & 1710   & ---  &  $7.67\pm 2.87$    &  $31.2\pm 68.0$   &  $19.1 \pm 41.1$  \\
                & 1900   & (c)  &  $10.56\pm 2.46$   &  $286.5\pm 104.7$ &  $8.3 \pm 231.3$  \\
                & 2090   & (c)  &  $1.73\pm 1.36$    &  $146.1\pm 52.9$  &  $20.5 \pm 348.9$  \\
   2.06         & 2100   & (c)  &  $3.12\pm 3.80$    &  $0.1\pm 355.9$   &  $295.6 \pm 86.3$  \\
 \hline
 \end{tabular}
  \end{center}
  \caption{The coupling constants of nucleon resonances to $\eta' N$ and relative phases extracted in the fit with including $P_{11}(1710)$ into the model. Only selected results are displayed, see the text for details.\\
  $^*$: The values are set to be zero in the fit.
  \label{Tab:1710plus}}
\end{table}

\begin{figure*}
  \begin{center}
{\includegraphics*[width=15.0cm]{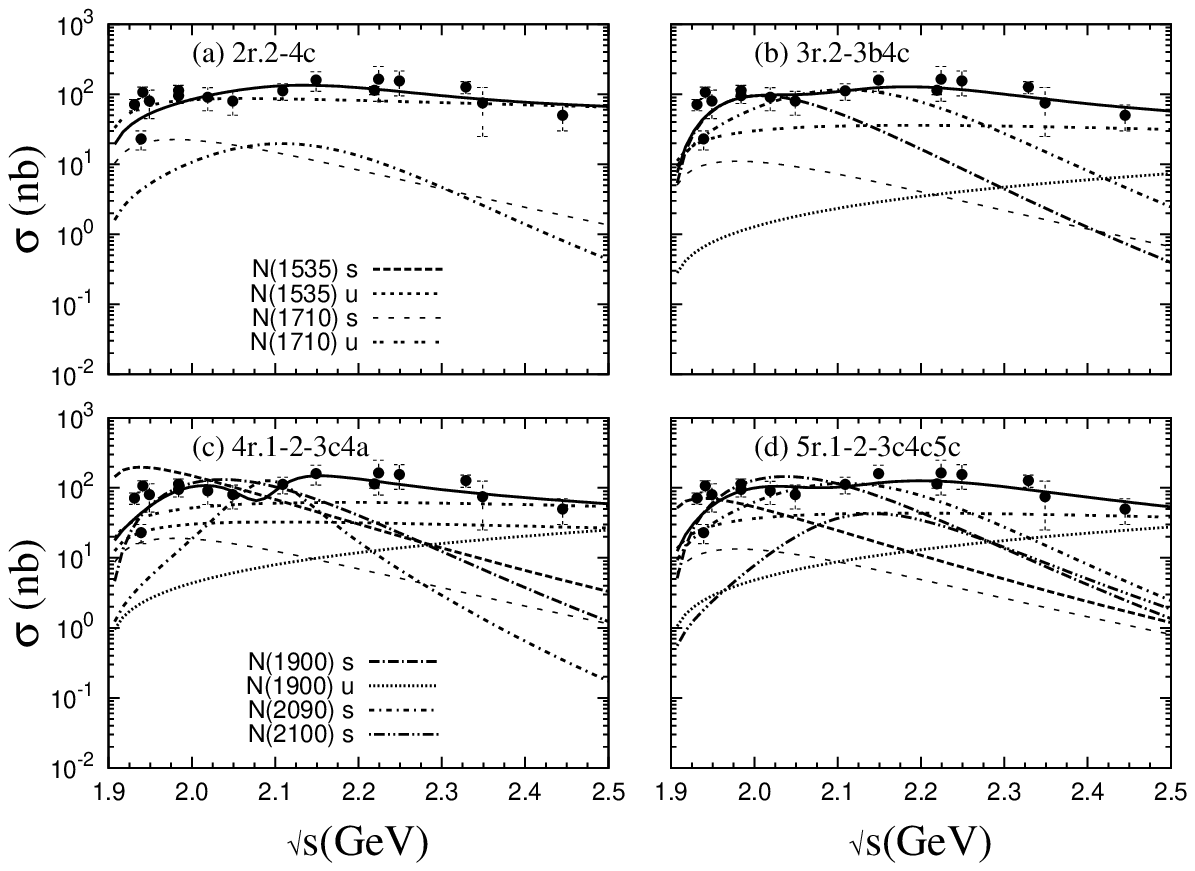}}
       \caption{
  Seleted diagrammatic representation for the fitted results with including $P_{11}(1710)$ into the model. The solutions are labeled as 2r.2-4c, 3r.2-3b4c, 4r.1-2-3c4a and 5r.1-2-3c4c5c, see Tab.~\ref{Tab:1710plus} for the reference of corresponding parameters.
  \label{Fig:1710plus}}
  \end{center}
\end{figure*}

\begin{figure*}
  \begin{center}
{\includegraphics*[width=15.0cm]{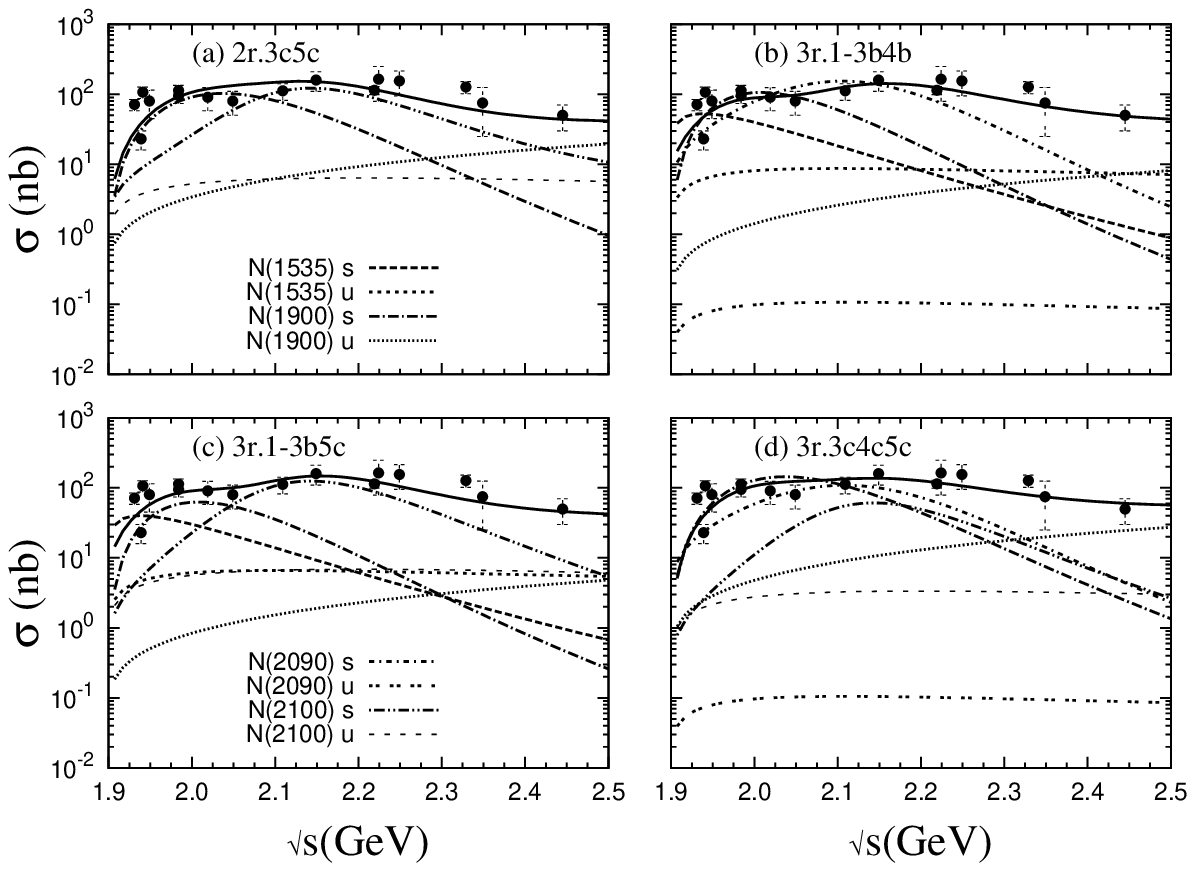}}
       \caption{
  Seleted diagrammatic representation for the fitted results with the combination of two resonances 2r.3c5c and three resonances 3r.1-3b4b, 3r.1-3b5c and 3r.3c4c5c, see Tab.~\ref{Tab:ccFitted2n3r} for the reference of corresponding parameters.
  \label{Fig:piN2n3r}}
  \end{center}
\end{figure*}

\begin{figure}
  \begin{center}
{\includegraphics*[width=8.0cm]{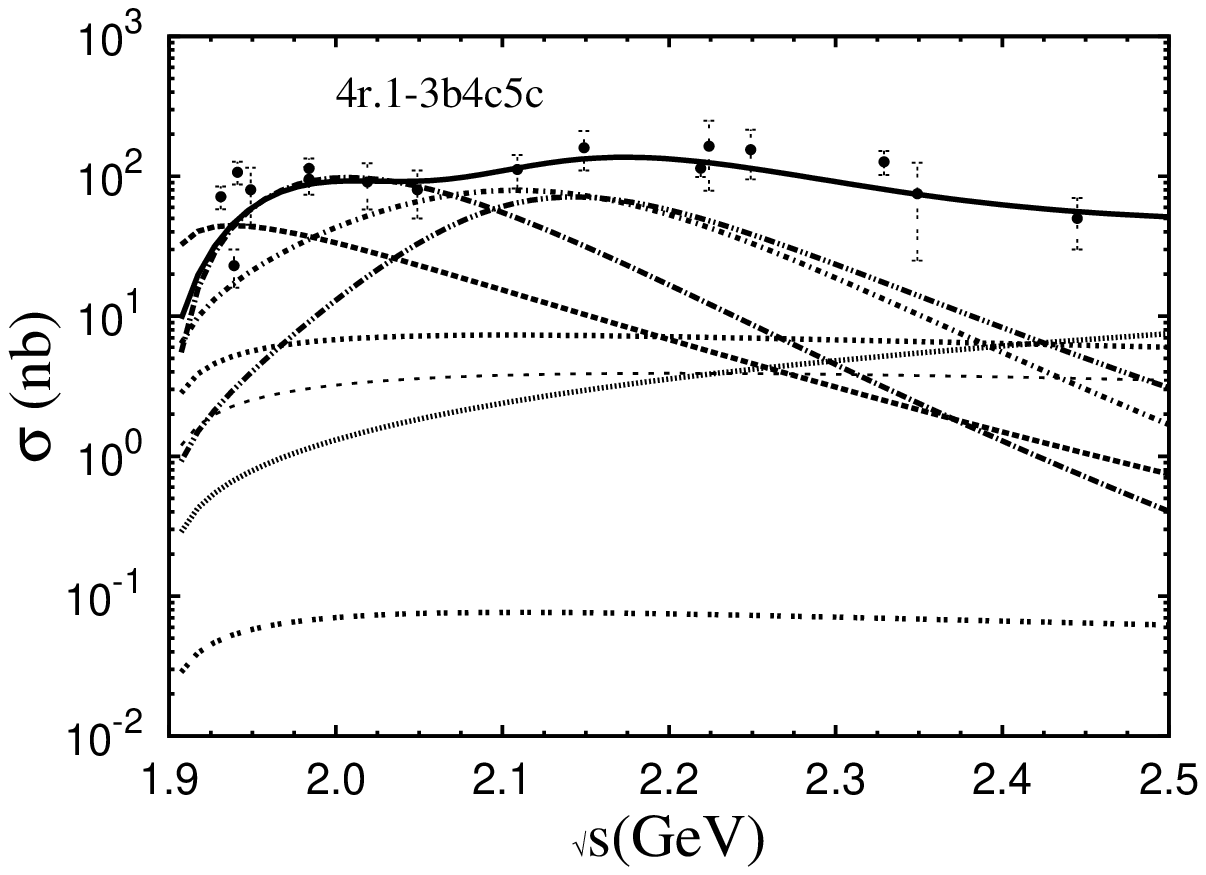}}
       \caption{
  Representative diagram for the fitted results with the combination of four resonances, e.g. 4r.1-3b4c5c, see Tab.~\ref{Tab:ccFitted4r} for the reference of corresponding parameters. The meaning of the lines is the same as those in Fig.~\ref{Fig:piN2n3r}.
  \label{Fig:piN4r}}
  \end{center}
\end{figure}

The total cross section of $\pi N \to \eta'N$ is around 100~$nb$ and roughly at the same level from the threshold to the center of mass (c.m.) energy $\sqrt{s} = 2.50$~GeV. The data also show an inconspicuous structure with two bumps at threshold and around 2.15~GeV, respectively. These prominent features are directly reflected in the following numerical fit results.

Evidently, one resonance alone can not give an excellent description of the data in this wide energy range. However, the $P_{11}(1710)$ could achieve a fair $\chi^2$ with the existing data, indicating as 1r.2- in Tab.~\ref{Tab:1710plus}. As demonstrated in Fig.~\ref{Fig:1710plus}, the $u$-channel $P_{11}(1710)$ diagram is significantly contributing a smooth background and the trend describes roughly the feature of the data without clear signal of generated resonances as mentioned above. At this stage it can not exclude this possibility due to the scarce data collection. Adding other resonances located above threshold, either wide or narrow, could improve the fit quality of the data. Due to so many solutions if including $P_{11}(1710)$ into the model, we only show some selected typical fitted parameters in Tab.~\ref{Tab:1710plus}, with the corresponding curves depicted in Fig.~\ref{Fig:1710plus}. We can see that the $\chi^2$ is getting a little better when the number of resonances is increasing. As shown in Fig.~\ref{Fig:1710plus}(c), the four-resonances solution gives a clear two bumps structure and is different from other solutions with relative plain curves. But the five-resonances solution 5r.1-2-3c4c5c is nearly in the same quality with the four-resonances one in terms of the $\chi^2$ values and its uncertainties of parameters is obviously bigger, exhausting the limitation of the data. This is also one of the reasons why we do not include more resonances in our model. It is a common feature that the $u$-channel $P_{11}(1710)$ contribution acts as an important smooth background and other resonances induces broad humps in all these solutions.

In the following text we would focus on the solution without the $P_{11}(1710)$ resonances. Two types of combination of two resonances have a good $\chi^2$. One is the $P_{11}(1710)$ with a resonances above threshold, with an representative example already shown as solution 2r.2-4c in Tab.~\ref{Tab:1710plus} and Fig.~\ref{Fig:1710plus}(a). The other one is the $P_{13}(1900)$ and broad $P_{11}(2100)$, and the extracted parameters are presented in the upper part of Tab.~\ref{Tab:ccFitted2n3r}, with a typical diagram of the total cross section of $\pi N \to \eta'N$ shown in Fig.~\ref{Fig:piN2n3r}(a). In the two-resonances solution, a relatively broad resonance above threshold is favored in order to reproduce the level behavior in the data. Here the $S_{11}(1535)$ seems to be not needed, but it should be cautious to draw this conclusion because the $S_{11}(1535)$ only contributes to the very close-to-threshold region, so adding only one resonance besides the $S_{11}(1535)$ can not explain well the big cross section at high energy. Surely, the $S_{11}(1535)$ alone could give a reasonable account of the very close-to-threshold data, as discussed in Ref.~\cite{caoetap2008} and shown in the three- and four-resonances solutions below.

Besides the three-resonances solution with the $P_{11}(1710)$, representatively shown in Fig.~\ref{Fig:1710plus}(b), all other three-resonances solutions with the $\chi^2$ around 2.3 are summarized in Tab.~\ref{Tab:ccFitted2n3r} and the typically diagrams are shown in Fig.~\ref{Fig:piN2n3r}(b-d). As can be seen in Fig.~\ref{Fig:piN2n3r}(b)(c), the $N^*(1535)$ could improve the $\chi^2$ from around 2.6 to about 2.3, mainly because of the better fit to the close-to-threshold data. However, its role could be substituted by other resonances above threshold and the combination of the $P_{13}(1900)$, $S_{11}(2090)$ and $P_{11}(2100)$ could give an equally good description of the data, as shown in Fig.~\ref{Fig:piN2n3r}(c). Obviously, we need more data to differentiate these mechanisms, especially those of the angular distributions and polarized observables.

About the four-resonances solutions, besides those mentioned above with including the $P_{11}(1710)$ (see Fig.~\ref{Fig:1710plus}(c) for representative example), the $S_{11}(1535)$, $P_{13}(1900)$, $S_{11}(2090)$ and $P_{11}(2100)$ together could reproduce the data, as listed in Tab.~\ref{Tab:ccFitted4r}. The $\chi^2$ ranges from 2.1 to 2.5 depending on the different widths of three resonances above threshold and the wider widths seem to be a little favored, mainly because of the feature of the data mentioned above. The contribution of $S_{11}(1535)$ is prominent at close-to-threshold region, as can be seen in Fig.~\ref{Fig:piN4r}. The $P_{13}(1900)$ is responsible for the first bump while the $S_{11}(2090)$ and $P_{11}(2100)$ together produce the second one. This is also happened in many other solutions when they are inclued.

It can be seen the situation is much more complicated than that in the $\pi^- p \to \phi n$ channel, where the $S_{11}(1535)$ resonance is dominant in a wide energy range~\cite{DaiPRD12}. But we can still find some common features in all the solutions besides those mentioned above. The contribution of the $u$-channel $S_{11}(1535)$ and $P_{11}(2100)$ is moderate at all energies and the $u$-channel $P_{11}(1900)$ is important at high energies. The $u$-channel $S_{11}(2090)$ term is very small and tends to be negligible. The interference effect can be seen, especially at threshold range, but it is not so important. This is understandable because we only fit the total cross section but the interference effect is more obvious in the differential cross sections and polarization observables. Moreover, it should be pointed out that the large errors of the relative phase $\phi_{u}$, most of which are compatible with zero degree, reflect the smallness of the corresponding $u-$channel contribution, but not only because of the limited data base.

Based on the above carefully analysis, we can conclude that at least one of the resonance among the $S_{11}(1535)$, $P_{11}(1710)$ and $P_{11}(1900)$ is required by the close-to-threshold data. We can also safely draw the conclusion that at least one relatively broad resonance above $\eta'N$ threshold is preferred. It is possible to further pin down the model parameters by the data of differential cross sections, which are however not at hand. It seems that the data of $\eta'$ photoproducion still do not give enough constraint to the model parameters, especially the masses and widths of the contributed resonances~\cite{HuangPRC13}. Anyway, the central values of extracted coupling constants are quite stable within the given uncertainties among these solutions, as shown in the tables. Especially, our present $g_{\eta'NN*(1535)}$ is consistent with the values in Ref.~\cite{caoetap2008}, and gives further support to the idea that the wave function of $N^*(1535)$ resonance has big $s\bar{s}$ component~\cite{AnSSG09,AnPRC11}. These reasons give us the confidence to use these extracted information from the $\pi N \to \eta'N$ reaction to study the $J/\psi \to p\bar{p}\eta'$ decay.

The BES Collaboration has accurately measured the BR($J/\psi \to p\bar{p}\eta'$) to be $(2.00\pm 0.23\pm 0.28) \times 10^{-4}$~\cite{Besnstar6}, about one order smaller than that of $J/\psi \to p\bar{p}\eta$ channel. With the total width $\Gamma_{J/\psi} = 92.9\pm 2.8$~keV~\cite{pdg2012}, we know the corresponding $\Gamma_{J/\psi \to p\bar{p}\eta'} = (1.86\pm 0.27\pm 0.32) \times 10^{-2}$~keV. The calculated  $\Gamma_{J/\psi \to p\bar{p}\eta'}$ using above parameters is in the range of 0.9 $\sim$ 8.2 $\times 10^{-2}$~keV, roughly compatible with the experiment within errors.

\begin{figure}
  \begin{center}
{\includegraphics*[width=9.cm]{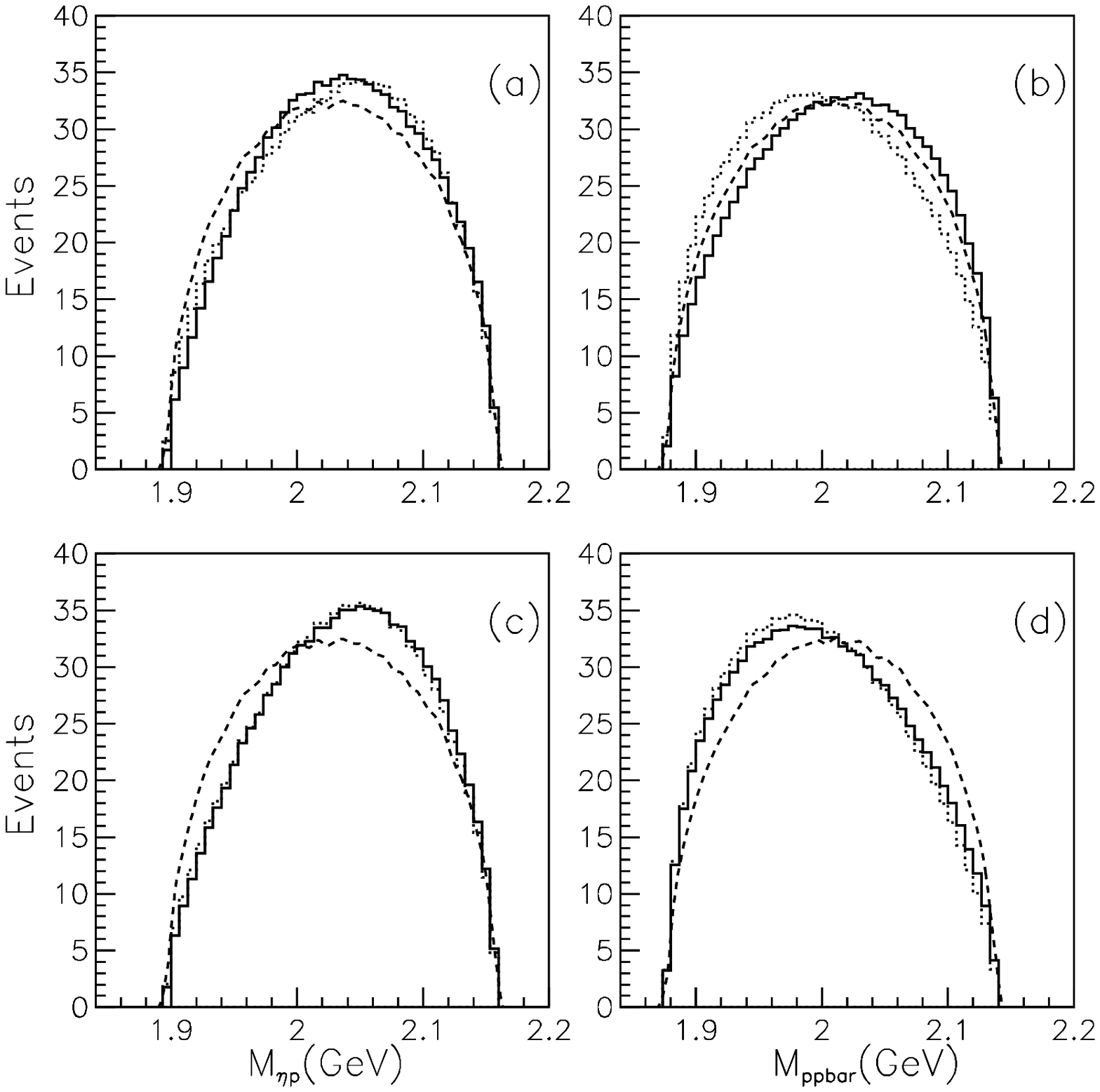}}
       \caption{
  The invariant mass spectrums for the $J/\psi \to p\bar{p}\eta'$ decay in the isobar model. The dashed curves are
  the pure phase-space distributions. The solid and dashed histogram are respectively for 3r.2-3b4c and 3r.3c4c5c in (a) and (b), while for 4r.1-3b4c5c and 5r.1-2-3c4c5c in (c) and (d).
      \label{Fig:JpsiDis}}
  \end{center}
\end{figure}

\begin{figure}
  \begin{center}
{\includegraphics*[width=9.cm]{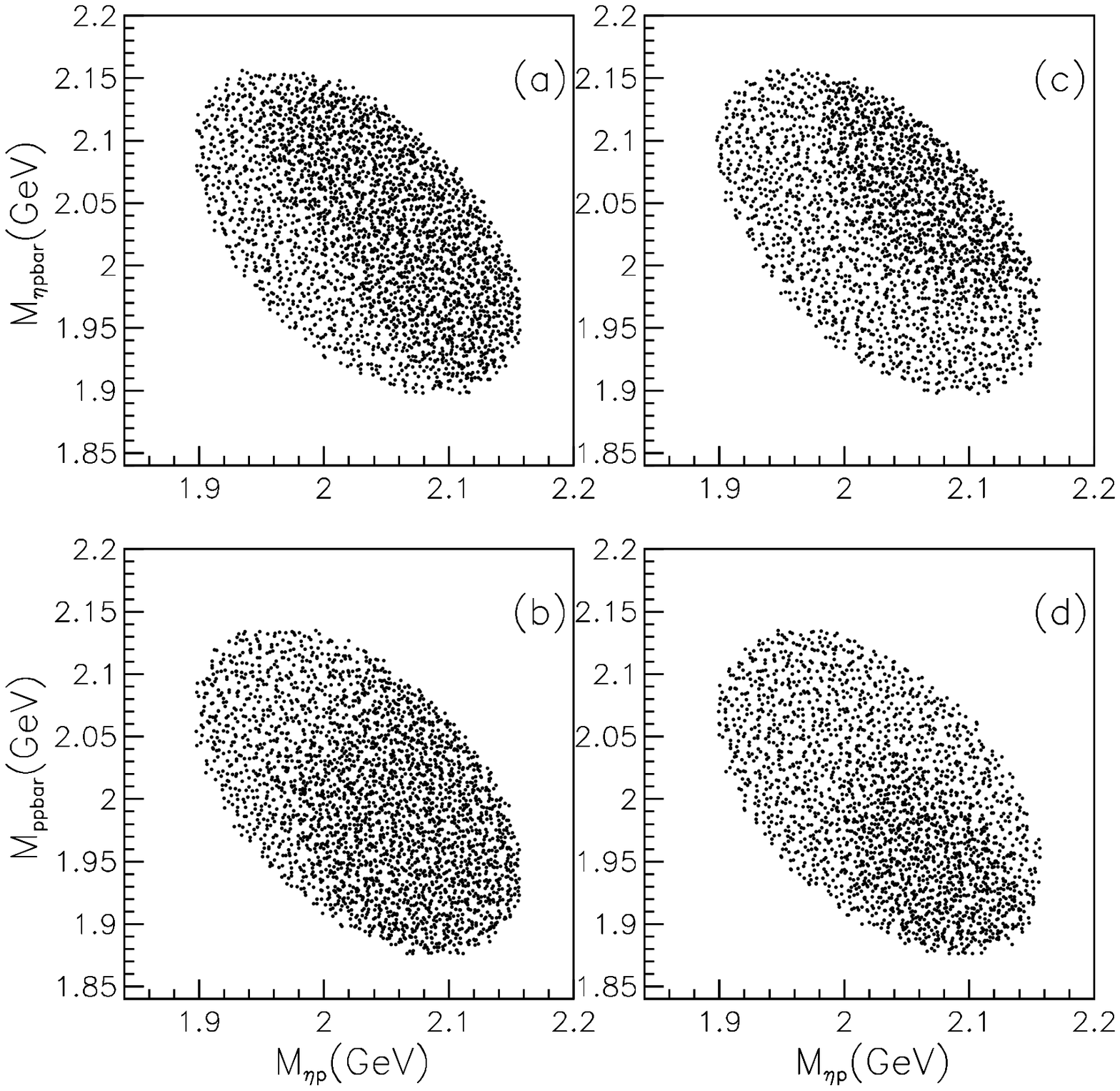}}
       \caption{
  The Dalitz plots for the $J/\psi \to p\bar{p}\eta'$ decay in the isobar model. The figure (a) and (b) are for 3r.3c4c5c, while (c) and (d) for 5r.1-2-3c4c5c.
      \label{Fig:JpsiDalitz}}
  \end{center}
\end{figure}

What we are more interested in is the invariant mass spectrums and Dalitz plots which may give us insight to the information of nucleon resonances. In Fig.~\ref{Fig:JpsiDis} we show the invariant mass spectrums of the solutions 3r.2-3b4c, 3r.3c4c5c, 4r.1-3b4c5c and 5r.1-2-3c4c5c as representative examples. It can be seen in Fig.~\ref{Fig:JpsiDis}(a) that the calculated spectrums do not have significant difference between solutions 3r.2-3b4c and 3r.3c4c5c, and this is also true for solutions 4r.1-3b4c5c and 5r.1-2-3c4c5c in Fig.~\ref{Fig:JpsiDis}(c)(d). In the $\eta'p$ spectrums, the enhancement is usually located above 2.0~GeV and the resonances below threshold move it a little closer to 2.0~GeV, as shown in Fig.~\ref{Fig:JpsiDis}(a). The more obvious effect of the resonances below threshold is appearing in the $p\bar{p}$ spectrums in Fig.~\ref{Fig:JpsiDis}(b). So in this situation it is probable to disentangle two $\eta'N$ production mechanism mentioned in the Introduction: the above or below threshold resonant contribution. Fig.~\ref{Fig:JpsiDalitz} depicts the Dalitz plots of the solutions 3r.3c4c5c and 5r.1-2-3c4c5c, which agrees with the conclusions in the invariant mass spectrums. As can be seen, the $p\bar{p}$-$\eta'p$ plots show more difference between various solutions so they are more suitable to study the $\eta'N$ production. Unfortunately, some of the invariant mass spectrums and Dalitz plots, e.g. solutions 2r.3a5c and 4r.1-3a4b5b, are very close to those of pure phase space, so they are unlikely to be distinguished from the totally non-resonant $\eta'N$ production mechanism. However, we could still expect that these two mechanisms would be recognized in the $\pi N \to \eta'N$ or $\gamma N \to \eta'N$ reaction. As a result, it is necessary to give a combined analysis to the data of various $\eta'N$ production channels in order to pin down the $\eta'N$ production mechanism.

It seems that it is more difficult to study the resonances in the $J/\psi \to p\bar{p}\eta'$ channel than in other decay channels, because all the above solutions do not show extraordinary resonances structures in the invariant mass spectrums and their deviation from phase space is not very significant. This is contrary to the case in the $J/\psi \to p\bar{p}\phi$~\cite{DaiPRD12}, the $J/\psi \to p \bar{p} \pi^0$~\cite{Besnstar5} and the $\psi(3686) \to p \bar{p} \pi^0$ decays~\cite{BESPRL2013}, where the resonance peaks obviously appear in the invariant mass spectrums and are also directly reflected in the Dalitz plots. Fortunately, the BESIII group is planning to collect around ten billion $J/\psi$ events in the near future, which is estimated to contain about two million events of $J/\psi \to p\bar{p}\eta'$ decay. Owing to such a large data base, it is still possible to study the nucleon resonances in the $J/\psi \to p\bar{p}\eta'$ decay.

\begin{table}[t]
  \begin{center}
  \begin{tabular}{cccccc}
 \hline
  Label/$\chi^2$   & $N^*$ & width & $g_{\eta'NR}$ & $\phi_{s}(^\circ)$ & $\phi_{u}(^\circ)$  \\
 \hline
  2r.3a5c & 1900   & (a)   &  $12.71\pm 2.07$  &  0.0$^*$          &  $209.7 \pm 101.1$  \\
   2.65   & 2100   & (c)   &  $5.48 \pm 0.72$  &  $151.3\pm 81.4$  &  $95.3 \pm 97.8$  \\
 \hline
  2r.3b5c & 1900   & (b)   &  $10.70\pm 1.76$  &  0.0$^*$          &  $169.4 \pm 86.2$  \\
   2.61   & 2100   & (c)   &  $5.56\pm 0.81$   &  $118.2\pm 72.8$  &  $44.9 \pm 84.4$  \\
 \hline
  2r.3c5c & 1900   & (c)   &  $8.88\pm 1.37$   &  0.0$^*$          &  $150.4 \pm 62.4$  \\
   2.63   & 2100   & (c)   &  $5.12\pm 1.05$   &  $98.5\pm 54.7$   &  $11.9 \pm 198.9$  \\
 \hline
 \hline
  3r.1-3b4b & 1535   & ---   &  $4.90\pm 2.38$   &  0.0$^*$          &  $55.1 \pm 59.4$  \\
            & 1900   & (b)   &  $13.02\pm 1.13$  &  $332.0\pm 58.5$  &  $242.9 \pm 65.5$  \\
   2.28     & 2090   & (b)   &  $1.37\pm 0.41$   &  $252.4\pm 24.5$  &  $57.2 \pm 104.7$  \\
 \hline
  3r.1-3c4b & 1535   & ---   &  $3.59\pm 1.47$   &  0.0$^*$          &  $247.5 \pm 42.8$  \\
            & 1900   & (c)   &  $9.36\pm 1.19$   &  $138.2\pm 28.2$  &  $47.0 \pm 38.6$  \\
   2.28     & 2090   & (b)   &  $1.58\pm 0.31$   &  $0.0\pm 13.1$    &  $245.1 \pm 89.9$  \\
 \hline
 \hline
  3r.1-3a5c & 1535   & ---   &  $4.83\pm 1.94$   &  0.0$^*$          &  $0.0 \pm 19.5$  \\
            & 1900   & (a)   &  $14.49\pm 2.93$  &  $265.9\pm 51.4$  &  $155.4 \pm 49.7$  \\
   2.38     & 2100   & (c)   &  $5.79\pm 0.99$   &  $87.0\pm 52.3$   &  $2.5 \pm 229.5$  \\
 \hline
  3r.1-3b5b & 1535   & ---   &  $5.36\pm 1.10$   &  0.0$^*$          &  $0.0 \pm 22.2$  \\
            & 1900   & (b)   &  $12.50\pm 1.77$  &  $264.1\pm 45.2$  &  $166.8 \pm 38.6$  \\
   2.43     & 2100   & (b)   &  $5.08\pm 1.00$   &  $122.6\pm 48.8$  &  $5.2 \pm 70.2$  \\
 \hline
  3r.1-3b5c & 1535   & ---   &  $4.27\pm 1.37$   &  0.0$^*$          &  $0.0 \pm 47.4$  \\
            & 1900   & (b)   &  $10.01\pm 1.61$  &  $0.0\pm 53.4$    &  $171.0 \pm 47.8$  \\
   2.35     & 2100   & (c)   &  $5.32\pm 1.32$   &  $140.9\pm 61.2$  & $4.0 \pm 216.0$  \\
 \hline
  3r.1-3c5b & 1535   & ---   &  $5.85\pm 1.30$   &  0.0$^*$          &  $0.0 \pm 55.9$  \\
            & 1900   & (c)   &  $8.94\pm 1.92$   &  $305.4\pm 62.3$  &  $162.5 \pm 56.3$  \\
   2.29     & 2100   & (b)   &  $4.13\pm 1.48$   &  $117.8\pm 49.1$  &  $12.4 \pm 208.1$  \\
 \hline
 \hline
  3r.3b4b5c & 1900   & (b)   &  $12.97\pm 1.05$  &  0.0$^*$          &  $230.6 \pm 67.3$  \\
            & 2090   & (b)   &  $0.91\pm 0.44$   &  $234.7\pm 36.9$  &  $70.3 \pm 116.4$  \\
   2.33     & 2100   & (c)   &  $4.91\pm 1.04$   &  $139.9\pm 62.6$  &  $91.9 \pm 59.4$  \\
 \hline
  3r.3b4c5c & 1900   & (b)   &  $13.04\pm 0.81$  &  0.0$^*$          &  $225.5 \pm 63.3$  \\
            & 2090   & (c)   &  $1.38\pm 0.44$   &  $235.7\pm 34.6$  &  $66.3 \pm 108.5$  \\
   2.31     & 2100   & (c)   &  $4.76\pm 1.03$   &  $130.9\pm 55.4$  &  $86.3 \pm 56.5$  \\
 \hline
  3r.3c4b5c & 1900   & (c)   &  $10.49\pm 0.89$  &  0.0$^*$          &  $241.4 \pm 53.6$  \\
            & 2090   & (b)   &  $1.15\pm 0.44$   &  $221.0\pm 26.0$  &  $90.3 \pm 117.3$  \\
   2.27     & 2100   & (c)   &  $4.33\pm 1.56$   &  $159.1\pm 61.0$  &  $108.9 \pm 58.6$  \\
 \hline
  3r.3c4c5c & 1900   & (c)   &  $10.53\pm 0.75$  &  0.0$^*$          &  $219.3 \pm 68.7$  \\
            & 2090   & (c)   &  $1.68\pm 0.75$   &  $219.8\pm 25.5$  &  $62.0 \pm 207.5$  \\
   2.29     & 2100   & (c)   &  $3.72\pm 1.77$   &  $141.5\pm 76.1$  &  $84.4 \pm 88.9$  \\
 \hline
 \end{tabular}
  \end{center}
  \caption{The coupling constants of nucleon resonances to $\eta' N$ and relative phases extracted in the fit with the combination of two or three resonances, mentioned as type I, see the text for details.\\
  $^*$: The values are set to be zero in the fit.
  \label{Tab:ccFitted2n3r}}
\end{table}

\begin{table}[t]
  \begin{center}
  \begin{tabular}{cccccc}
 \hline
  Label/$\chi^2$   & $N^*$ & width & $g_{\eta'NR}$ & $\phi_{s}(^\circ)$ & $\phi_{u}(^\circ)$  \\
 \hline
  4r.1-3a4b5b & 1535   & ---   &  $5.54 \pm  2.88$  &  0.0$^*$          & $0.0 \pm 71.8$  \\
              & 1900   & (a)   &  $14.88 \pm 1.67$  &  $0.0 \pm 36.1$   & $182.8 \pm 47.7$  \\
              & 2090   & (b)   &  $1.11 \pm  0.49$  &  $251.2 \pm 46.7$ & $0.0 \pm 349.1$  \\
   2.50       & 2100   & (b)   &  $4.21 \pm  2.34$  &  $78.5 \pm 50.6$  & $0.0 \pm 36.2$  \\
 \hline
  4r.1-3a4b5c & 1535   & ---   &  $2.63 \pm 1.09$   &  0.0$^*$          & $160.5 \pm 63.2$  \\
              & 1900   & (a)   &  $15.17 \pm 2.45$  &  $8.0 \pm 29.4$   & $333.3 \pm 76.5$  \\
              & 2090   & (b)   &  $1.28 \pm 0.42$   &  $292.6 \pm 23.7$ & $147.5 \pm 114.2$  \\
   2.46       & 2100   & (c)   &  $3.66 \pm 2.01$   &  $325.1 \pm 33.7$ &  $157.0 \pm 46.0$  \\
 \hline
  4r.1-3a4c5c & 1535   & ---   &  $3.36 \pm 1.48$   & 0.0$^*$           &  $163.2 \pm 76.1$  \\
              & 1900   & (a)   &  $14.39 \pm 2.60$  & $323.3 \pm 59.6$  &  $0.0 \pm 50.2$  \\
              & 2090   & (c)   &  $1.90 \pm 0.74$   & $283.3 \pm 35.8$  &  $158.1 \pm 121.0$  \\
   2.38       & 2100   & (c)   &  $3.09 \pm 2.85$   & $357.8 \pm 358.5$ &  $134.8 \pm 85.7$  \\
 \hline
  4r.1-3b4b5b & 1535   & ---   &  $3.78 \pm 2.58$   &  0.0$^*$          &  $52.5 \pm 92.1$  \\
              & 1900   & (b)   &  $13.17 \pm 1.13$  &  $0.0 \pm 240.6$  &  $239.4 \pm 98.3$  \\
              & 2090   & (b)   &  $1.29 \pm 0.58$   &  $252.9 \pm 30.0$ &  $54.8 \pm 281.3$  \\
   2.23       & 2100   & (b)   &  $2.42 \pm 2.81$   &  $188.3 \pm 258.0$&  $65.2 \pm 101.9$  \\
 \hline
  4r.1-3b4b5c & 1535   & ---   &  $5.55 \pm 4.67$   &  0.0$^*$          &  $347.7 \pm 53.2$  \\
              & 1900   & (b)   &  $12.0 \pm 1.14$   &  $286.6 \pm 65.3$ &  $194.9 \pm 61.5$  \\
              & 2090   & (b)   &  $1.05 \pm 0.51$   &  $210.7 \pm 73.1$ &  $346.5 \pm 61.9$  \\
   2.39       & 2100   & (c)   &  $3.19 \pm 3.29$   &  $9.2 \pm 20.2$   &  $348.3 \pm 5.5$  \\
 \hline
  4r.1-3b4c5b & 1535   & ---   &  $5.62 \pm 3.16$   &  0.0$^*$          &  $0.0 \pm 47.4$  \\
              & 1900   & (b)   &  $12.50 \pm 1.34$  &  $0.0 \pm 61.4$   &  $196.2 \pm 63.1$  \\
              & 2090   & (c)   &  $1.68 \pm 0.66$   &  $251.7 \pm 37.5$ &  $19.7 \pm 226.7$  \\
   2.19       & 2100   & (b)   &  $3.34 \pm 2.81$   &  $91.8 \pm 77.4$  &  $36.2 \pm 70.0$  \\
 \hline
  4r.1-3b4c5c & 1535   & ---   &  $4.50 \pm 3.75$   &  0.0$^*$          &  $0.0 \pm 38.7$  \\
              & 1900   & (b)   &  $12.51 \pm 1.55$  &  $0.0 \pm 70.3$   &  $189.4 \pm 73.9$  \\
              & 2090   & (c)   &  $1.43 \pm 0.96$   &  $240.5 \pm 62.4$ &  $13.6 \pm 281.4$  \\
   2.17       & 2100   & (c)   &  $4.01 \pm 3.17$   &  $92.4 \pm 86.1$  &  $31.3 \pm 65.9$  \\
 \hline
  4r.1-3c4c5c & 1535   & ---   &  $3.62 \pm 2.78$   &  0.0$^*$          &  $0.0 \pm 271.5$  \\
              & 1900   & (c)   &  $9.76 \pm 1.59$   &  $0.0 \pm 305.4$  &  $196.9 \pm 133.8$  \\
              & 2090   & (c)   &  $1.44 \pm 1.33$   &  $226.6 \pm 50.4$ &  $88.7 \pm 214.3$  \\
   2.12       & 2100   & (c)   &  $3.88 \pm 2.77$   &  $123.1 \pm 94.3$ &  $99.3 \pm 97.5$  \\
 \hline
 \end{tabular}
  \end{center}
  \caption{The coupling constants of nucleon resonances to $\eta' N$ and relative phases extracted in the fit with the combination of four resonances, see the text for details.\\
  $^*$: The values are set to be zero in the fit.
  \label{Tab:ccFitted4r}}
\end{table}

\section{Summary} \label{sec:summary}

In a short summary, we perform a full analysis in the effective Lagrangian approach to extract the information from the data of $\pi N \to \eta'N$ reaction. Though the present data do not restrict enough the production mechanism, we find that either a subthreshold resonance or a broad $P$-wave state at near threshold seems to be required and at least one broad resonance above $\eta'N$ threshold is preferred by the data. This is useful to our understanding of the nucleon resonances coupling strongly to the $\eta'N$ channel. From the present analysis of the data, we could calculate the $J/\psi \to p\bar{p}\eta'$ decay and find that there is no so distinct resonances structures in the invariant mass spectrums as in other decay channels. However, the $J/\psi \to p\bar{p}\eta'$ decay may still be useful for discriminating the below or above resonant contribution in the $\eta'N$ production. The BESIII group are encouraged to do the PWA on the basis of future large data base. The data of $\pi N \to \eta'N$ and $\gamma N \to \eta'N$ reactions are also suggested to be included to give a combined analysis in order to finally determine the $\eta'N$ production mechanism and resonances contribution.

It is worth to point out that our results are also meaningful to the study of the $p\bar{p} \to J/\psi \eta'$ at PANDA@FAIR~\cite{QYLinppbar}. We can speculate that the contribution of the $t-$channel nucleon resonances should be much more important than that of the $t-$channel nucleon pole in this reaction. The similar conclusion should be also applied to the associate production of charmonium state and other non-strange meson (except for $\pi$-meson) in $p\bar{p}$ annihilation, namely $p\bar{p} \to J/\psi \eta$, $p\bar{p} \to J/\psi \omega$ and $p\bar{p} \to J/\psi \phi$. As a result, the total cross section of these reactions are expected to be enhanced so they could be more easily measured at PANDA.

\begin{acknowledgments}

We would like to thank Dr. J. P. Dai for useful discussion. This work was supported by the National Natural Science Foundation of China (Grant Nos. 11347146, 11405222, 11275235 and 11175220).

\end{acknowledgments}


\begin{thebibliography}{1}
%
\bibitem{pdg2012}J. Beringer {\it et al.} (Particle Data Group), Phys. Rev. D \textbf{86}, 010001 (2012); K. A. Olive {\it et al.} (Particle Data Group), Chin. Phys. C \textbf{38}, 090001 (2014).
\bibitem{BoCh11}A. V. Anisovich, E. Klempt, and V. A. Nikonov et al., Eur. Phys. J. \textbf{A47}, 27 (2011).
\bibitem{BoCh12}A. V. Anisovich, R. Beck, E. Klempt et al., Eur. Phys. J. \textbf{A48}, 15 (2012).
\bibitem{BoCh08}V. A. Nikonov, A. V. Anisovich, and E. Klempt et al., Phys. Lett. \textbf{B662}, 245 (2008).
\bibitem{BoCh12qtet}A. V. Anisovich, E. Klempt, and V. A. Nikonov et al., Phys. Lett. \textbf{B711}, 167 (2012).
\bibitem{Feuster98}T. Feuster and U. Mosel, Phys. Rev. C {\bf 58}, 457 (1998); {\it ibid.} {\bf 59}, 460 (1999).
\bibitem{Penner2002}G. Penner and U. Mosel, Phys. Rev. C {\bf 66}, 055211 (2002); {\it ibid.} {\bf 66}, 055212 (2002).
\bibitem{Shklyar04J}V. Shklyar, G. Penner, and U. Mosel, Eur. Phys. J. {\bf A21}, 445 (2004).
\bibitem{Shklyar05ome}V. Shklyar, H. Lenske, U. Mosel, and G. Penner, Phys. Rev. C {\bf 71}, 055206 (2005).
\bibitem{Shklyar05lam}V. Shklyar, H. Lenske, and U. Mosel, Phys. Rev. C {\bf 72}, 015210 (2005).
\bibitem{Shklyar07eta}V. Shklyar, H. Lenske, and U. Mosel, Phys. Lett. {\bf B650}, 172 (2007).
\bibitem{Shklyar12eta}V. Shklyar, H. Lenske, and U. Mosel, Phys.Rev. C {\bf 87}, 015201 (2013).
\bibitem{caoKSigma}X. Cao, V. Shklyar, and H. Lenske, Phys. Rev. C {\bf 88}, 055204 (2013).
\bibitem{Manley1992}D. M. Manley and E. M. Saleski, Phys. Rev. D \textbf{45}, 4002 (1992).
\bibitem{Manley2012}M. Shrestha and D. M. Manley, Phys. Rev. C \textbf{86}, 055203 (2012).
\bibitem{Arndt2012}R. L. Workman, W. J. Briscoe, M. W. Paris, and I. I. Strakovsky, Phys. Rev. C \textbf{85}, 025201 (2012).
\bibitem{Doring11}M. D\"{o}ring, C. Hanhart, and F. Huang {\it et al.}, Nucl. Phys. \textbf{ A851}, 58 (2011).
\bibitem{Huang12}F. Huang, M. D\"{o}ring, H. Haberzettl {\it et al.}, Phys.Rev. C \textbf{85}, 054003 (2012).
\bibitem{Doring12}D. R\"{o}nchen, M. D\"{o}ring, F. Huang {\it et al.}, Eur. Phys. J. \textbf{A49}, 44 (2013).
\bibitem{LeePhysRe}A. Matsuyama, T. Sato, and T.-S. H. Lee, Phys. Rep. {\bf 439}, 193 (2007).
\bibitem{KamanoCC}H. Kamano, S. X. Nakamura, T.-S. H. Lee, and T. Sato, Phys. Rev. C {\bf 88}, 035209 (2013).
\bibitem{Kamano2pi}H. Kamano, Phys. Rev. C {\bf 88}, 045203 (2013).
\bibitem{BCLiuPRL06}B.-C. Liu and B.-S. Zou, Phys. Rev. Lett. {\bf 96} 042002 (2006).
\bibitem{JJXiePLB07}J.-J. Xie and B.-S. Zou, Phys. Lett. {\bf B649}, 405 (2007).
\bibitem{JJXiePRC08}J.-J. Xie, B.-S. Zou, and H.-C. Chiang, Phys. Rev. C {\bf 77}, 015206 (2008).
\bibitem{JJXiePRC10}J.-J. Xie and C. Wilkin, Phys. Rev. C {\bf 82}, 025210 (2010).
\bibitem{JJXiePRC11}J.-J. Xie, H.-X. Chen, and E. Oset, Phys. Rev. C {\bf 84}, 034004 (2011).
\bibitem{JJXiePRC13}J.-J. Xie and B.-C. Liu, Phys.Rev. C {\bf 87}, 045210 (2013).
\bibitem{caoCPL2008}X. Cao, X.-G. Lee, and Q.-W. Wang, Chin. Phys. Lett. {\bf 25}, 888 (2008).
\bibitem{caoetap2008}X. Cao and X.-G. Lee, Phys. Rev. C {\bf 78}, 035207 (2008).
\bibitem{caophi2009}X. Cao, J.-J. Xie, B.-S. Zou, and H.-S. Xu, Phys. Rev. C {\bf 80}, 025203 (2009).
\bibitem{caoCPC2009}X. Cao, Chin. Phys. C {\bf 33}, 1381 (2009).
\bibitem{caotwopi}X. Cao, B.-S. Zou, and H.-S. Xu, Phys. Rev. C {\bf 81}, 065201 (2010).
\bibitem{caoNPA2011}X. Cao, B.-S. Zou, and H.-S. Xu, Nucl. Phys. {\bf A861}, 23  (2011).
\bibitem{caoIJMPA}X. Cao, B.-S. Zou, and H.-S. Xu, Int. J. Mod. Phys. A {\bf 26}, 505 (2011).
\bibitem{Besnstar5}M. Ablikim {\it et al.} (BES Collaboration), Phys. Rev. D {\bf 80}, 052004 (2009).
\bibitem{BESRoper06}M. Ablikim {\it et al.} (BES Collaboration), Phys. Rev. Lett. {\bf 97}, 062001 (2006).
\bibitem{BESPRL2013}M. Ablikim {\it et al.} (BES Collaboration), Phys. Rev. Lett. {\bf 110}, 022001 (2013).
\bibitem{BESPRD06}M. Ablikim {\it et al.} (BES Collaboration), Phys. Rev. D {\bf 74}, 012004 (2006).
\bibitem{Besnstar2}J. Z. Bai {\it et al.} (BES Collaboration), Phys. Lett. {\bf B510}, 75 (2001);
\bibitem{Besnstar3}H. X. Yang {\it et al.} (BES Collaboration), Int. J. Mod. Phys. {\bf A20}, 1985 (2005);
\bibitem{Besnstar4}M. Ablikim {\it et al.} (BES Collaboration), Phys. Lett. {\bf B659}, 789 (2008).
\bibitem{Besnstar6}M. Ablikim {\it et al.} (BES Collaboration), Phys. Lett. {\bf B676}, 25 (2009).
\bibitem{Besnstar7}M. Ablikim {\it et al.} (BES Collaboration), Phys. Rev. D {\bf 87}, 112004 (2013).
\bibitem{CLEO2005}R. A. Briere {\it et al.} (CLEO Collaboration), Phys. Rev. Lett. {\bf 95}, 062001 (2005).
\bibitem{CLEO2007}S. B. Athar {\it et al.} (CLEO Collaboration), Phys. Rev. D {\bf 75}, 032002 (2007).
\bibitem{ZouEPJA2003}B.-S. Zou and D. V. Bugg, Eur. Phys. J. {\bf A16}, 537 (2003).
\bibitem{ZouPRC2003}B.-S. Zou and F. Hussain, Phys. Rev. C {\bf 67}, 015204 (2003).
\bibitem{ZouEPJA2005}S. Dulat and B.-S. Zou, Eur. Phys. J. {\bf A26}, 125 (2005).
\bibitem{Torres13}K. P. Khemchandani, A. Mart\'{i}nez Torres, H. Nagahiro, and A. Hosaka, Phys. Rev. D {\bf 88}, 114016 (2013).
\bibitem{OsetPLB11}E. Oset and A. Ramos, Phys. Lett. {\bf B704}, 334 (2011).
\bibitem{Nakappeta1}K. Nakayama, J. Speth, and T. -S. H. Lee, Phys. Rev. C {\bf 65}, 045210 (2002).
\bibitem{Nakappeta2}K. Nakayama, J. Haidenbauer, C. Hanhart, and J. Speth, Phys. Rev. C {\bf68}, 045201 (2003).
\bibitem{HuangPRC13}F. Huang, H. Haberzett, and K. Nakayama, Phys. Rev. C {\bf 87}, 054004 (2013).
\bibitem{WHLiangJPG02}W.-H. Liang, P.-N. Shen, J.-X. Wang, and B.-S. Zou, J. Phys. {\bf G28}, 333 (2002).
\bibitem{WHLiangEPJA04}W.-H. Liang, P.-N. Shen, B.-S. Zou4, and A. Faessler, Eur. Phys. J. {\bf A21}, 487 (2004).
\bibitem{BarnesPRD10}T. Barnes, X. Li, and W. Roberts, Phys. Rev. D {\bf 81}, 034025 (2010).
\bibitem{Okubo1984}R. Sinha and S. Okubo, Phys. Rev. D {\bf 30}, 2333 (1984).
\bibitem{DaiPRD12}J.-P. Dai, P.-N. Shen, J.-J. Xie, and B.-S. Zou, Phys. Rev. D {\bf 85}, 014011 (2012).
\bibitem{JShiPRD11}J. Shi, J.-P. Dai, and B.-S. Zou, Phys. Rev. D {\bf 84}, 017502 (2011).
\bibitem{AnSSG09}C.-S. An and B.-S. Zou, Sci. Sin. G {\bf 52}, 1452 (2009).
\bibitem{AnPRC11}C.-S. An and B. Saghai, Phys. Rev. C {\bf 84}, 045204 (2011).
\bibitem{QYLinppbar}Q.-Y Lin, H.-S. Xu, and X. Liu, Phys. Rev. D {\bf 86}, 034007 (2012).

\end{thebibliography}
\end{document}